\documentclass{elsart}

\usepackage{color}

\usepackage{cite}

 \usepackage{graphicx}

\usepackage{amssymb}
\usepackage{multirow}

\usepackage{color}

\begin{document}

\begin{frontmatter}



\title{Theoretical estimation of metabolic network robustness against multiple reaction knockouts using branching process approximation}


\author[KIT,corr]{Kazuhiro Takemoto}
\ead{takemoto@bio.kyutech.ac.jp}
\author[BIC]{Takeyuki Tamura}
\author[BIC]{Tatsuya Akutsu}

\address[KIT]{Department of Bioscience and Bioinformatics, Kyushu Institute of Technology, Kawazu 680-4, Iizuka, Fukuoka 820-8502, Japan}

\address[BIC]{Bioinformatics Center, Institute for Chemical Research, Kyoto University, Gokasho, Uji, Kyoto, 611-0011, Japan}

\corauth[corr]{Corresponding author.}

\begin{abstract}
In our previous study, we showed that the branching process approximation is useful for estimating metabolic robustness, measured using the {\it impact degree}.
By applying a theory of random family forests, we here extend the branching process approximation to consider the knockout of {\it multiple} reactions, inspired by the importance of multiple knockouts reported by recent computational and experimental studies.
In addition, we propose a better definition of the number of offspring of each reaction node, allowing for an improved estimation of the impact degree distribution obtained as a result of a single knockout. Importantly, our proposed approach is also applicable to multiple knockouts.
The comparisons between theoretical predictions and numerical results using real-world metabolic networks demonstrate the validity of the modeling based on random family forests for estimating the impact degree distributions resulting from the knockout of multiple reactions.
\end{abstract}


\end{frontmatter}

\section{Introduction}
\label{sec:intro}
Robustness is an important concept for understanding how living organisms
adapt to changing environments, as well as for understanding how they are able to survive when carrying mutated genes  \cite{Stelling2004,kitano07}.
In particular, metabolic robustness is of increasing interest not only to researchers in the field of basic biology, but also to those in biotechnology and medical research because metabolic processes are essential for physiological functions, and are responsible for maintaining life.

The development of high-throughput methods has facilitated the collection and compilation of large metabolic network datasets, which are stored in databases such as the Kyoto Encyclopedia of Genes and Genomes (KEGG) \cite{kanehisa10} and the Encyclopedia of Metabolic Pathways (MetaCyc) \cite{keseler11}.
In recent years, numerous methods and measures have been developed for analyzing metabolic robustness in the context of gene/reaction knockout by using available metabolic network data.

Many of these methods and measures are based on \emph{flux balance analysis} (FBA).
Edwards and Palsson used the change in the optimal objective
function value (e.g., growth rate) as a measure of robustness
against the change in a particular reaction \cite{edwards00}.
Segr\`{e} et al. \cite{segre02} proposed the minimization of metabolic adjustment (MOMA)
method, which predicts the flux vectors by minimizing the Euclidean distance
between the mutant and wild type.
Deutscher et al. \cite{deutscher08} proposed another measure by combining FBA with
the Shapley value in game theory.
With respect to FBA,
\emph{elementary flux modes} (EFMs) have also been used to
analyze the robustness of metabolic networks,
where an EFM is a minimal set of reactions that can operate at the steady state
\cite{papin04}.
Wilhelm et al. \cite{wilhelm04} proposed a measure based on the numbers of EFMs
before and after knockout,
which was later extended to include the knockout of multiple reactions \cite{behre08}.
In order to evaluate robustness for the production of
a specific target compound(s),
several studies have used
a minimum reaction cut based on FBA and/or EFM
\cite{acuna09,burgard03,haus08,klamt04},
which involves a minimum set of reactions (or enzymes),
the removal of which leads to prevention of the production of a specific
set of compounds.
Other approaches based on FBA/EFM have also been implemented \cite{larhlimia11}.

Boolean modeling is an alternative way to model metabolic networks, whereby
the activity of each reaction or compound is represented
by either 0 (inactive) or 1 (active) and
reactions and compounds are modeled as AND nodes and OR nodes respectively.
Handorf et al. \cite{handorf05} introduced the concept of {\em scope} based on Boolean
modeling and applied it to analyses of the robustness of metabolic networks.
Li et al. \cite{li09}, Sridhar et al. \cite{sridhar08}, and Tamura et al. \cite{tamura09} developed
integer programming-based methods for determining the minimum reaction cut
under Boolean models.
Lemke et al. \cite{lemke04} defined the \emph{damage} as the number of reactions
inactivated by the knockout of a single reaction under a Boolean model.
Smart et al. \cite{smart08} refined the concept of damage
by introducing the \emph{topological flux balance} (TFB) criterion.
Jiang et al. defined the \emph{impact degree} as
the number of reactions inactivated by knockout of
a specified reaction \cite{jiang09} under a Boolean model.
Although there are some differences in the treatment of reversible reactions,
the damage and the impact degree are very similar concepts.

To date, most studies have focused on the prediction and/or accuracy of
robustness measures but have given less consideration to the distribution
of such measures.
Lemke et al. \cite{lemke04} analyzed the distribution of damage using computer simulation.
Smart et al. \cite{smart08} performed a similar analysis.
In addition, they applied percolation theory and branching processes
to the analysis of the distribution of cluster sizes of damaged subnetworks \cite{smart08}; however, they did not explicitly estimate damage distribution (i.e., impact degree distribution).

Until recently, theoretical frameworks for estimating the tolerance of metabolic networks to various failures were poorly established.
Motivated by this, in our previous study \cite{Takemoto2012}, we analyzed the distribution of impact degree triggered by random knockout of a single reaction using a branching process theory \cite{lee04,saichev05}. By treating the propagation of the impact triggered by the knockout of a reaction as a branching process approximation, the relevance of which had been shown in the context of loading-dependent cascading failure \cite{Dobson2004,Dobson2005,Kim2010}, we demonstrated that the branching process model (or theory) reflects the observed impact degree distributions.
In addition, Lee et al. \cite{lee12} also recently demonstrated the use of a Boolean model and a theory of branching process in this context.

As above, most previous studies focused on the impact of a single knockout.
In recent years, however, multiple-knockout experiments have been actively performed, and have shown new interesting results on metabolic robustness, such as synergetic effects resulting from multiple knockouts \cite{Deutscher2006,Nakahigashi2009,Suthers2009}.
Therefore, computational and theoretical frameworks need to be extended to include multiple knockouts.
For example, Deutscher et al. \cite{deutscher08} discussed the impact of multiple knockouts in yeast metabolism based on the Shapley value from game theory.
Tamura et al. \cite{tamura11} proposed an efficient method for computing metabolic robustness in the context of impact degree.
However, theoretical approaches remain incomplete.

In this study, by extending the branching process approximation proposed in our previous study \cite{Takemoto2012}, we show that the branching process approximation (specifically, the assumption of a random family forest, which is a collection of family trees) is also useful for estimating the distribution of the impact degree triggered by the random knockout (or disruption) of {\it multiple} reactions.

\section{Impact Degree}

Here, we briefly review the impact degree \cite{jiang09} and
its extensions \cite{tamura09,tamura11}.
The \emph{impact degree} was originally proposed by Jiang et al. as
a measure of the importance of each reaction in a metabolic network
\cite{jiang09},
and is defined as the number of inactivated reactions caused
by the knockout of a single reaction.
Since the effect of cycles was not considered in their study,
Tamura et al. extended the impact degree so that the effect of cycles is
taken into account by introducing the
maximal valid assignment concept \cite{tamura09}.
Furthermore, they extended it to cope with the knockout of multiple reactions
\cite{tamura11}.

Let $V_c = \{C_1,\ldots,C_m\}$ and $V_r = \{R_1,\ldots,R_n\}$ be
a set of \emph{compound nodes} and a set of \emph{reaction nodes} respectively,
where $V_c \cap V_r = \{\}$.
A \emph{metabolic network} is defined as a bipartite directed graph
$G(V_c \cup V_r,E)$
in which each edge is directed either from a node in $V_c$ to a node in
$V_r$, or from a node in $V_r$ to a node in $V_c$.
Each of the included reactions and compounds takes 1 of 2 states: 0 (inactive) or
1 (active).

The impact degree for the knockout of multiple reactions is computed as follows
\cite{tamura11}.
Let $V_{ko} = \{R_{i_1},\ldots,R_{i_D}\}$ be a set of
reactions that have been knocked out.
We start with the global state, such that all compounds are active
(i.e., $C_i=1$ for all $C_i \in V_c$) and all reactions except for those in
$V_{ko}$ are active
(i.e., $R_i=1$ for all $R_i \in V_r - V_{ko}$ and $R_i=0$
for all $R_i \in V_{ko}$).
Then, we update the states of reactions and compounds using the following rules.
\begin{enumerate}
\item A reaction is inactivated if any predecessor (i.e., substrate) or
successor (i.e., product) is inactive.
\item A compound is inactivated if all predecessors or all
successors are inactive.
\end{enumerate}
We repeat this procedure until reaching a stable global state,
which is determined uniquely regardless of the order of updates \cite{tamura09}.
The impact degree for $V_{ko}$ is the number of inactive reactions
(i.e., reactions with value 0) in the stable global state.
Although this procedure simultaneously gives the definition and
algorithm for the impact degree,
Tamura et al. developed a much more efficient algorithm to compute it
\cite{tamura11}.

\begin{figure}[th]
\begin{center}
\includegraphics{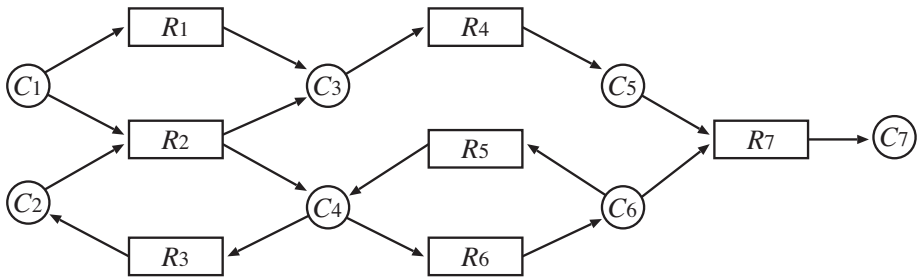}
\caption{Example of a metabolic network.
Circles and boxes correspond to compounds and reactions,
respectively.}
\label{exnet}
\end{center}
\end{figure}

Fig.~\ref{exnet} shows an example of a metabolic network.
If $R_1$ is knocked out, the other reactions remain active and thus
the impact degree is 1.
If $R_2$ is knocked out, $C_2$ is inactivated and then $R_3$ is inactivated.
However, $C_3$ and $C_4$ remain active, and thus the impact degree is 2.
If both $R_1$ and $R_2$ are knocked out,
$C_3$ is inactivated, and $R_4$ and $R_7$ are inactivated.
Since $R_1,R_2,R_3,R_4,R_7$ are inactive in the stable global state,
the resulting impact degree is 5.

\section{Branching process approximation}
\label{sec:branch}
We have extended the branching process approximation, previously reported in \cite{Takemoto2012}, to include the estimation of the impact degree distributions obtained as a result of the knockout of multiple reactions.

The branching process \cite{Harris1989} is a stochastic process in which each progenitor generates offspring according to a fixed probability distribution called the offspring distribution.
Since the propagation of an impact on a network is essentially similar to cascading failures, which is a sequence of failures caused by an accident,
branching process approximation is useful for estimating impact degree distributions, although it requires the assumption of a network tree structure (see Sec. \ref{sec:discuss} for details of model limitations).

\subsection{The number of offspring in metabolic networks}
\label{sec:offspring}
We need to define the number of offspring for each reaction node in metabolic networks in order to analyze the impact degree distributions using the branching process approximation.

To obtain the number of offspring, in the previous study \cite{Takemoto2012}, we used the reaction network obtained as the unipartite projection of the metabolic network, where we draw an edge from reaction A to reaction B when at least 1 product of A is a substrate of B (see Sec 3.1 in \cite{Takemoto2012} for details). Assuming that the impact spreads though reactions whose substrates are synthesized via unique metabolic reactions (i.e., reaction nodes with the indegree of 1) when assuming tree structures of networks, we defined the number $d_i$ of offspring for reaction node $i$ in metabolic networks as follows: $d_i = k^{\mathrm{out}}_i$ if $k^{\mathrm{in}}_i=1$ and $0$ otherwise, where $k^{\mathrm{out}}_i$ and $k^{\mathrm{in}}_i$ are the outdegree and indegree of reaction node $i$ in the reaction network, respectively.

\begin{figure}[th]
\begin{center}
\includegraphics{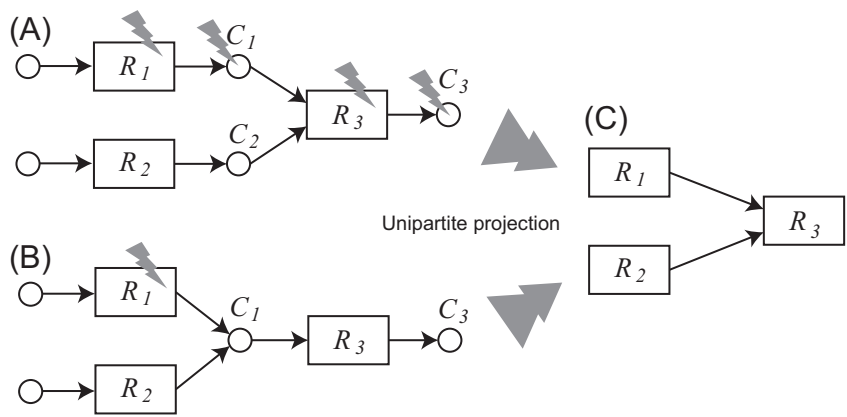}
\caption{Example illustrating how different metabolic networks (A and B) are converted to the same reaction network (C) through the unipartite projection.
Circles and boxes correspond to compounds and reactions, respectively.
Lightning bolts indicate the impacts.
}
\label{fig2:reaction_net}
\end{center}
\end{figure}

However, this definition is not appropriate when different metabolic networks (Figs. \ref{fig2:reaction_net}A and \ref{fig2:reaction_net}B) are projected into the same reaction network (Fig. \ref{fig2:reaction_net}C).
In this case, the similar offspring distribution (or potential of spreading) is defined between these different metabolic networks although the tendency of impact spreading is clearly different between the networks.
For example, in particular, we consider the knockout of the reaction $R_1$ (i.e., the inactivation of $R_1$) in Fig. \ref{fig2:reaction_net}.
In Fig. \ref{fig2:reaction_net}A, the reaction $R_3$ is also inactive (i.e., the impact spreads) after the inactivation of the reaction $R_1$ because it requires the compound $C_1$ synthesized through the reaction $R_1$.
On the other hand, in Fig. \ref{fig2:reaction_net}B, the impact does not spread because the compound $C_1$ required by the reaction $R_3$ can be synthesized through the reaction $R_2$ even if the reaction $R_1$ is inactive.

To distinguish between these cases, we consider {\it spreading edges} that contributes to the impact spreading on reaction networks.
In particular, spreading edges are defined as directed edges between a reaction node pair (i.e., source and target nodes) whose interjacent chemical compounds are synthesized by the source reaction only.
For example, in Fig. \ref{fig2:reaction_net}A, the interjacent compound (i.e., $C_1$) between the reaction nodes $R_1$ and $R_3$ is generated through the reaction $R_1$ only; thus, the directed edge from the reaction node $R_1$ to the reaction node $R_3$ in Fig. \ref{fig2:reaction_net}C is defined as a spreading edge.
On the other hand, in Fig. \ref{fig2:reaction_net}B, the interjacent compound $C_1$ is obtained either through the reactions $R_1$ or $R_2$; thus, the directed edge from the reaction node $R_1$ to the reaction node $R_3$ in Fig. \ref{fig2:reaction_net}C is not regarded as a spreading edge.

After finding the spreading edges in reaction networks according to this definition, the number of offspring of a reaction node is defined as the number of out-going spreading edges.
For example, the number of offspring of the reaction node $R_1$ and $R_2$ in Fig. \ref{fig2:reaction_net}A are both 1; however, they are both 0 in Fig. \ref{fig2:reaction_net}B.
In this manner, we can determine the difference in the tendency of impact spreading between these metabolic networks.

We finally obtain the offspring distribution $P(d)$ from an empirical metabolic network as
\begin{equation}
P(d)=\frac{1}{N}\sum_{i=1}^{N}\delta(d,d_i),
\label{eq:off_dist}
\end{equation}
where $N$ corresponds to the total number of reaction nodes.
The function $\delta(x,y)$ is the Kronecker's delta function, and it returns $1$ if $x=y$ and 0 otherwise.

\subsection{Branching process models}
Here, we explain the estimation method for the impact degree distribution obtained as a result of the knockout of either a single reaction or multiple reactions, based on a theory of branching processes.

\subsubsection{Case I: knockout of a single reaction}
We previously investigated the impacts of a single knockout \cite{Takemoto2012}; here, we briefly summarize these methods and findings as a basis for extending our approach to include cases in which multiple reactions are knocked out.

In this study, we consider a branching process model (i.e., the empirical model in Ref. \cite{Takemoto2012}) using empirical offspring distributions, obtained using Eq. (\ref{eq:off_dist}).
Because the impact degree can be regarded as the total number of offspring through a branching process \cite{Takemoto2012}, we obtained the distribution of the total number of offspring to estimate the impact degree distribution.
However, the branching process model we described previously does not count the first impact (i.e., progenitor) although counting the first impact is necessary for estimating the impact degree distribution obtained as a result of the knockout of multiple reactions.
Thus, we estimate the distribution of the impact degree, including the first impact, triggered by the knockout of a single reaction, as follows.

Let $F(s)$ be the probability generating function of the impact degree $r$ (i.e., the total number of offspring, including the progenitor); the function $F(s)$ satisfies the recursive relation \cite{Otter1949,Harris1989,Pitman1998}:
\begin{equation}
F(s)=sf(F(s)),
\end{equation}
where $f(s)$ denotes the probability generating function of the number $d$ of offspring of each reaction node:
\begin{equation}
f(s)=\sum_{d=0}^{d_{\max}}P(d)s^d,
\end{equation}
where $d_{\max}$ is the maximum number of offspring.

Using the formula reported by Burman and Lagrange and the relation of
\begin{equation}
P(r)=\left.\frac{1}{r!}\frac{d^rF(s)}{ds^r}\right|_{s=0},
\end{equation}
the distribution $P(r)$ (i.e., impact degree distribution) is derived from the above implicit equation as the following explicit equation \cite{Otter1949}:
\begin{equation}
P(r)=\left.\frac{1}{r!}\left[\frac{d^{r-1}}{ds^{r-1}}(f(s))^r\right]\right|_{s=0}.
\label{eq:P(r)_emp}
\end{equation}
The impact degree distributions obtained as a result of the random knockout of a single reaction are estimated through this equation.

\subsubsection{Case II: knockout of multiple reactions}
\label{sec:multi_esti}
In this study, we assume the impact spreading triggered by the knockout of multiple reactions is represented by the independent branching process, given an initial number of $k$ individuals (progenitors; i.e., the random family forest, which is a collection of $k$ family trees) \cite{Pitman1998}.

Because the branching process with $k$ initial progenitors can be defined as $k$ independent copies of a branching process with a single initial progenitor \cite{Pitman1998}, the impact degree triggered by the knockout of multiple reactions can be described as the sum of the total number of offspring through a branching process with a single initial knockout.
Thus, the probabilistic generating function $F_k(s)$ of the impact degree, which is induced by the knockout of $k$ reactions, is derived using the probabilistic generating function $F(s)$ of the impact degree that results from a single knockout. This is demonstrated as follows:
\begin{equation}
F_k(s)=[F(s)]^k.
\label{eq:P(r)_k}
\end{equation}

To obtain the probabilistic generating function $F(s)$ of the impact degree triggered by a single knockout, we consider 2 cases.
In the first case, we use an estimated value for $F(s)$ that is derived using Eq. (\ref{eq:P(r)_emp}):
\begin{equation}
F(s)=\sum_{r=0}^{r_{\max}}P(r)s^r.
\label{eq:F(s)_2}
\end{equation}
In the second case, we use an empirical value of $F(s)$ (i.e., the probabilistic generating function calculated according to Eq. (\ref{eq:F(s)_2}), using the empirical $P(r)$ computed by numerical simulations).
Using the empirical $F(s)$, we can purely evaluate the validity of the assumption of random family forests.
$r_{\max}$ corresponds to the maximum impact degree in the empirical $P(r)$.

Finally, the impact degree distribution $P_k(r)$ obtained as a result of the knockout of $k$ reactions is estimated as
\begin{equation}
P_k(r)=\mathrm{the \ coefficient \ of \ } s^r \mathrm{\ in \ } F_k(s).
\label{eq:P_k(r)}
\end{equation}

\section{Evaluation of the branching process approximation}
We evaluated the efficiency of the above estimation methods for the impact degree distributions using real-world metabolic networks of several species.

We focused on 2 bacteria, (i.e., {\it Escherichia coli} and {\it Bacillus subtilis}), and 2 eukaryotes, (i.e., {\it Saccharomyces cerevisiae} (yeast) and {\it Homo sapiens} (human)), whose metabolic pathways have been well characterized using experimental approaches in a previous study \cite{Takemoto2012}. We downloaded metabolic network data for each species from the KEGG database \cite{kanehisa10,KEGG}. Each dataset was represented as bipartite networks, as shown in Fig. \ref{exnet},.

For these species, the impact degree distributions obtained as a result of the knockout of $k$ reactions were numerically calculated using an efficient algorithm \cite{tamura11}.
In this study, we considered cases of $k=1$ (i.e., single knockout) and $k=2$ and 3 (i.e., multiple knockouts).
We compared these observed impact degree distributions using the theoretical distributions that were obtained as follows.

\begin{figure}[tbp]
\begin{center}
	\includegraphics{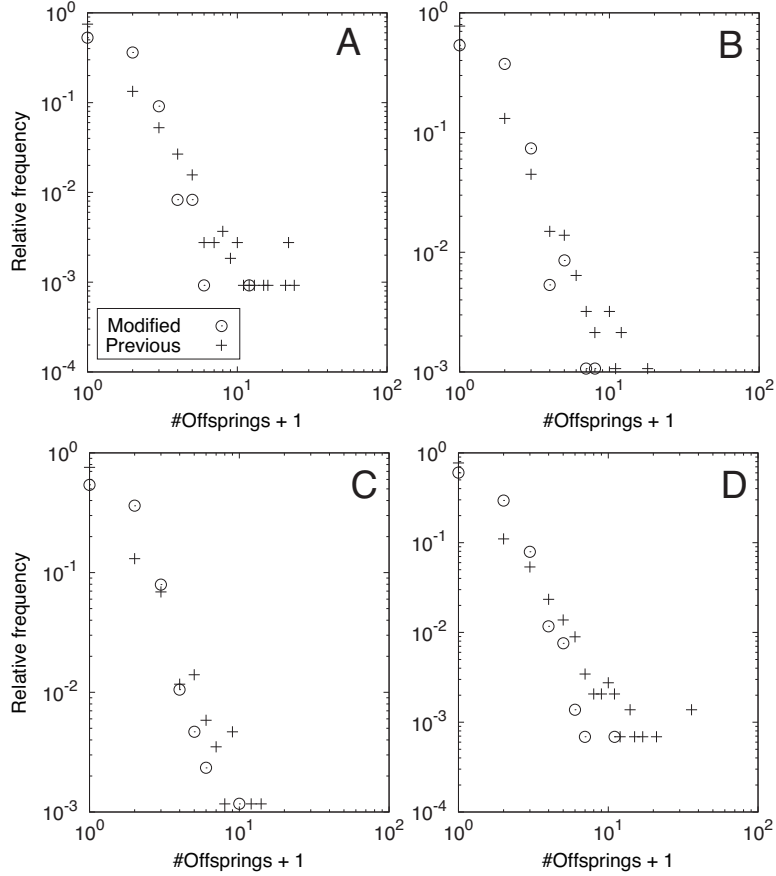}
	\caption{
	The offspring distributions of
	{\it Escherichia coli} (A),
	{\it Bacillus subtilis} (B),
	{\it Saccharomyces cerevisiae} (C), and
	{\it Homo sapiens} (D).
	The open circles and crosses indicate the distributions obtained using the modified and previous definitions of the number of offspring, respectively.
	}
	\label{fig:fig3_offspring}
\end{center}
\end{figure}

According to Sec. \ref{sec:offspring}, we constructed the reaction networks through the unipartite projection of bipartite metabolic networks from the KEGG database and obtained the offspring distributions from the reaction networks (Fig. \ref{fig:fig3_offspring}).
The number of offspring follows a power-law-like distribution.
This may be because of the heterogeneous (or scale-free) connectivity in metabolic networks \cite{Barabasi2004}.
The offspring distributions are slightly different between the modified definition based on spreading edges and the previous definition in Ref. \cite{Takemoto2012}.
Table \ref{table:parameter} summarizes the parameters extracted from the reaction networks, and it shows that the mean number of offspring, an important parameter for descriptions of branching processes, is either overestimated or underestimated using the previously described definition, which is in contrast to the use of our modified definition presented here.
The cases of {\it E. coli} and {\it H. sapiens} represent $\mu_{\mathrm{previous}}>\mu_{\mathrm{modified}}$, whereas the cases of {\it B. subtilis} and {\it S. cerevisiae} show $\mu_{\mathrm{previous}}<\mu_{\mathrm{modified}}$.
These differences in offspring distributions affect the accuracy of predicting the impact degree distributions (Table \ref{table:accuracy}).
Note that the number of nodes and the mean number of offspring are slightly different between the previous study \cite{Takemoto2012} and this study because of the removal of redundant metabolic reactions in the database.

\begin{table}[tbp]
\caption{Parameters extracted from real-world reaction networks. The character \# indicates ``the number of".
$\mu_{\mathrm{previous}}$ and $\mu_{\mathrm{modified}}$ are the mean number of offspring of a reaction node estimated using our previously described definition \cite{Takemoto2012} and the modified definition described here based on spreading edges (this study), respectively.
}
\label{table:parameter}
\begin{center}
\begin{tabular}{l|cccc}
\hline
\hline
Species & \#Nodes & \#Directed edges & $\mu_{\mathrm{previous}}$ & $\mu_{\mathrm{modified}}$ \\
\hline
{\it Escherichia coli} & 1085 & 3824 & 0.63 & 0.62 \\
{\it Bacillus subtilis} & 937 & 2799 & 0.47 &  0.58 \\
{\it Saccharomyces cerevisiae} & 856 & 2328 &  0.48 & 0.59 \\
{\it Homo sapiens} & 1425 & 5514 & 0.59 & 0.54 \\
\hline
\hline
\end{tabular}
\end{center}
\end{table}

\subsection{Case I: knockout of a single reaction}
\begin{figure}[tbp]
\begin{center}
	\includegraphics{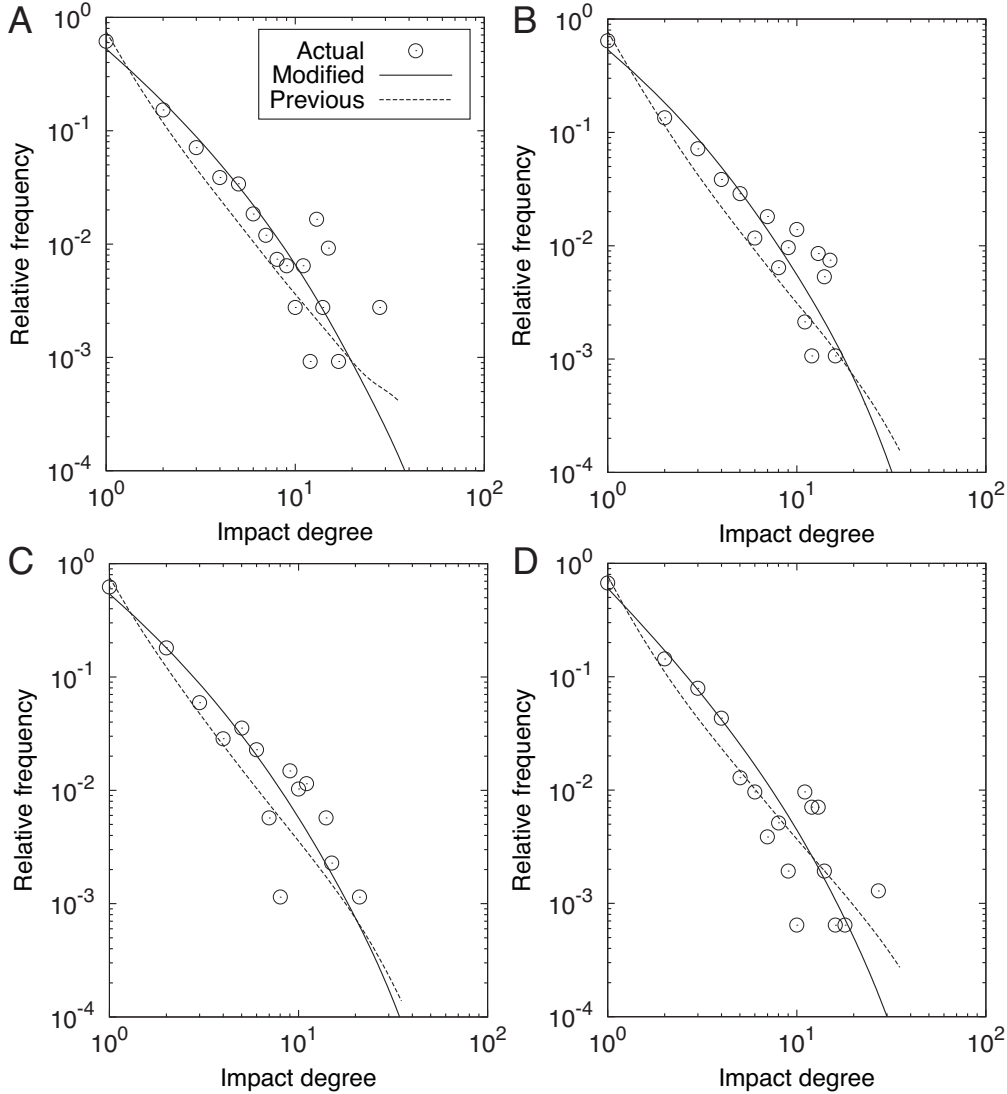}
	\caption{
	The impact degree distributions obtained after the knockout of a single reaction in
	{\it Escherichia coli} (A),
	{\it Bacillus subtilis} (B),
	{\it Saccharomyces cerevisiae} (C), and
	{\it Homo sapiens} (D).
	The circles indicate observed data.
	The solid lines and dashed lines correspond to the theoretical distributions estimated by Eq. (\ref{eq:P(r)_emp}) using the modified and previous definitions of offspring distributions, respectively.
	}
	\label{fig:fig3_single_knockout}
\end{center}
\end{figure}

As reported in our previous study \cite{Takemoto2012}, the branching process approximation is useful for estimating the impact degree distributions (Fig. \ref{fig:fig3_single_knockout}).
Because of the subcritical case (i.e., $\mu<1$ as shown in Table \ref{table:parameter}), the impact degree distributions do not follow a clear power law, and they show an exponential cut-off for larger impact degrees.
Assuming a Poisson distribution with the mean of $\mu$, for example, the total number of offspring (i.e., impact degree) $r$ is distributed according to the Borel distribution \cite{Dobson2004}: $P(r)=(\mu r)^{r-1}e^{-\mu r}/r!$.
Using Stirling's formula (i.e., $r!\approx \sqrt{2\pi r}r^r e^{-r}$), the above equation leads to the approximation
\begin{equation}
P(r)\propto r^{-3/2}e^{-r(\ln \mu-\mu+1)}.
\end{equation}
When $\mu=1$ (i.e., the critical case), the impact degree follows the power-law distribution with the exponent of $-3/2$ when assuming a Poisson offspring distribution \cite{Otter1949,Harris1989,Dobson2004,Kim2010}.
Moreover, assuming a power-law offspring distribution (i.e., $P(d) \propto 1/d^{\gamma+1}$, where $\gamma<2$), the impact degree follows $P(r)\propto 1/r^{1+1/\gamma}$ \cite{saichev05}.

To evaluate the goodness of fit between the empirical distributions and theoretical distributions from Eq. (\ref{eq:P(r)_emp}), we performed the Kolmogorov-Smirnov (KS) test (Table \ref{table:accuracy}).
The KS test shows that the theoretical distributions estimated using the branching process approximation are in good agreement with empirical distributions of real-world metabolic networks and that the estimation using the modified definition of the number of offspring is better than that using the previous definition.
In particular, the poor agreement between the impact degree distributions of the {\it H. sapiens} metabolic network calculated using the theoretical estimation and real data, as reported in the previous study \cite{Takemoto2012}, was improved as a result of the consideration of spreading edges (see Sec. \ref{sec:offspring}).
Note that the KS distance and the $P$-value are different between the previous study \cite{Takemoto2012} and this study because the previous study does not consider the first impact when calculating the impact degree.

\begin{table}[tbhp]
\caption{Comparison of prediction accuracy (i.e., goodness of fit) for the impact degree distributions between the previous definition and the modified definition of $d_i$: Kolmogorov-Smirnov (KS) distance, defined as $\sup_{x}|R(x)-M(x)|$, where $R(x)$ and $M(x)$ are empirical distributions and theoretical distributions, respectively.
The parenthetic values indicate the logarithmic $P$-values $p$ from the KS test, defined as $-\log_{10}(p)$.
A smaller KS distance and logarithmic $P$-values indicate a higher goodness of fit between the empirical distribution and theoretical distribution.
The highlighted values correspond to the best accuracy.}
\label{table:accuracy}
\begin{center}
\begin{tabular}{l|cc}
\hline
\hline
Species & Previous version & Modified version\\
\hline
{\it Escherichia coli} & 0.12 (1.01) & {\bf 0.08 (0.54)}\\
{\it Bacillus subtilis} & 0.18 (1.76) & {\bf 0.12 (1.27)}\\
{\it Saccharomyces cerevisiae} & 0.15 (1.13) & {\bf 0.13 (1.09)}  \\
{\it Homo sapiens} & 0.18 (3.33)  & {\bf 0.10 (1.33)}\\
\hline
\hline
\end{tabular}
\end{center}
\end{table}

\subsection{Case II: knockout of multiple reactions}
\begin{figure}[tbp]
\begin{center}
	\includegraphics{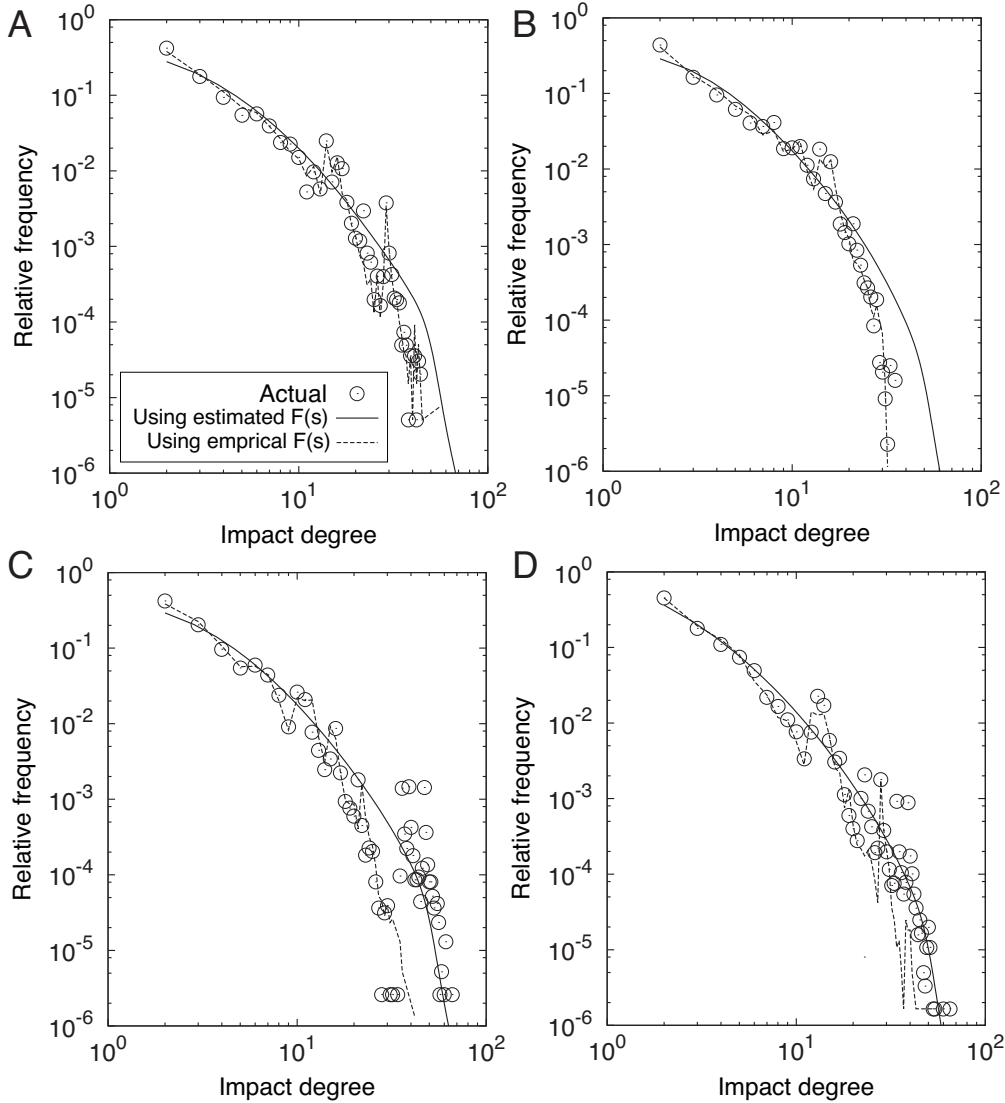}
	\caption{
	The impact degree distributions obtained following the knockout of 2 reactions in
	{\it Escherichia coli} (A),
	{\it Bacillus subtilis} (B),
	{\it Saccharomyces cerevisiae} (C), and
	{\it Homo sapiens} (D).
	The circles indicate observed data.
	The solid lines and dashed lines correspond to the theoretical distributions from Eq. (\ref{eq:P_k(r)}) using the estimated $F(s)$ and empirical $F(s)$, respectively.
	}
	\label{fig:fig4_double_knockout}
\end{center}
\end{figure}

\begin{figure}[tbp]
\begin{center}
	\includegraphics{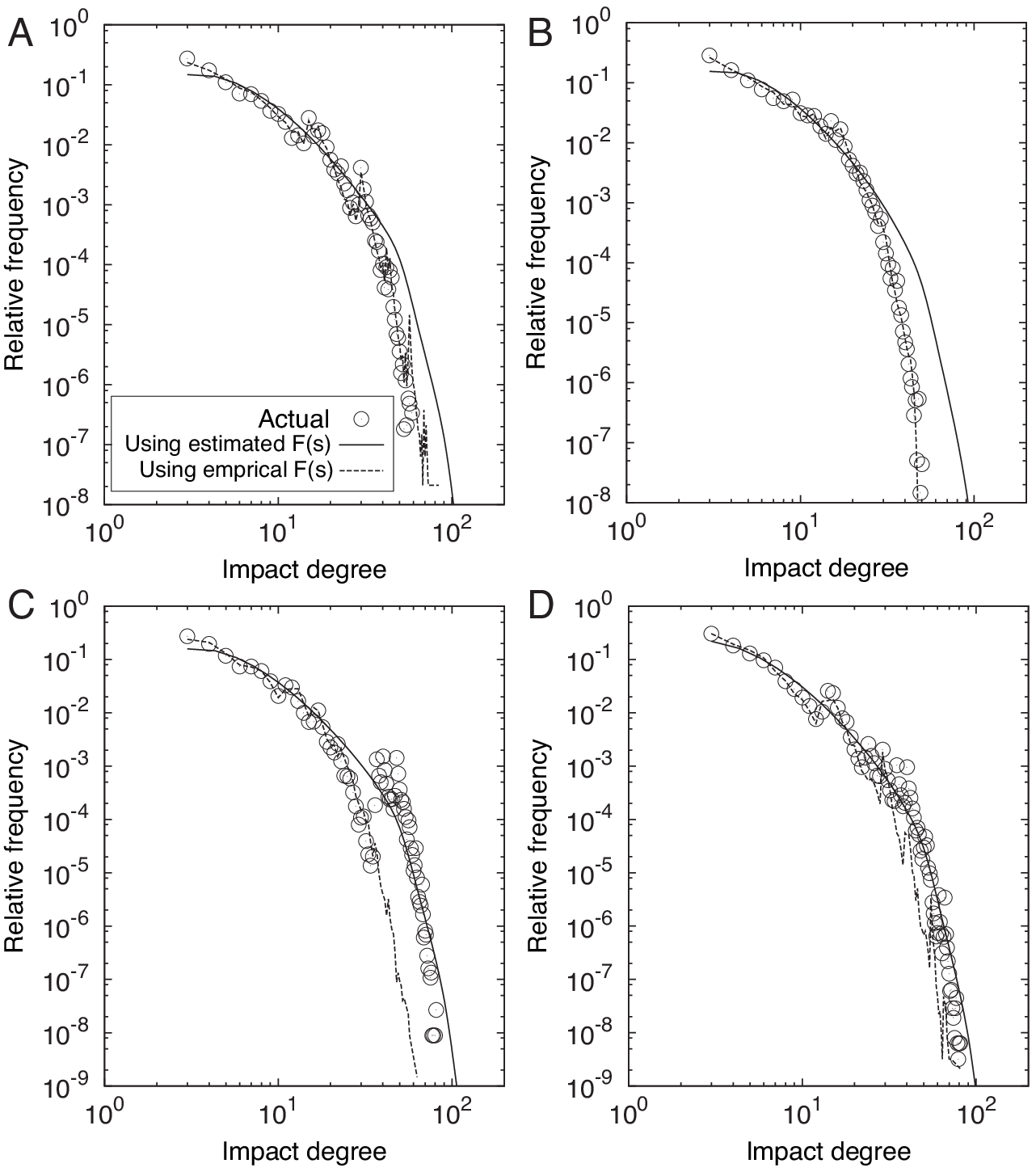}
	\caption{
	The impact degree distributions obtained as a result of the knockout of 3 reactions of
	{\it Escherichia coli} (A),
	{\it Bacillus subtilis} (B),
	{\it Saccharomyces cerevisiae} (C), and
	{\it Homo sapiens} (D).
	The circles indicate observed data.
	The solid lines and dashed lines correspond to the theoretical distributions from Eq. (\ref{eq:P_k(r)}) using the estimated $F(s)$ and empirical $F(s)$, respectively.
	}
	\label{fig:fig5_triple_knockout}
\end{center}
\end{figure}

We evaluated the efficiency of the branching process approximation when knocking out 2 reactions (Fig. \ref{fig:fig4_double_knockout}) and when knocking out 3 reactions (Fig. \ref{fig:fig5_triple_knockout}).
As explained in Sec. \ref{sec:multi_esti}, we consider both the estimated $F(s)$ and empirical $F(s)$ when estimating the impact degree distributions resulting from the knockout of multiple reactions.
We used the definition based on spreading edges when calculating the estimated $F(s)$ due to its improved accuracy (Table \ref{table:accuracy}).

Overall, the branching process approximation (i.e., the modeling based on random family forests \cite{Pitman1998}) is also useful for estimating the impact degree distribution for the knockout of multiple reactions, despite some exceptions.

The larger impact degrees of the theoretical distribution calculated from {\it B. subtilis} (Figs. \ref{fig:fig4_double_knockout}B and \ref{fig:fig5_triple_knockout}B) using the estimated $F(s)$ and the observed distributions are not similar.
This weaker fit is caused by the prediction accuracy in the branching process approximation (i.e., Eq. (\ref{eq:P(r)_emp})) rather than the assumption of random family forests, illustrated by the fact that the theoretical distributions derived using the empirical $F(s)$ are in good agreement with the observed distributions.

In the case of {\it S. cerevisiae} (Figs. \ref{fig:fig4_double_knockout}C and \ref{fig:fig5_triple_knockout}C), our data suggest that the assumption of random family forests is less suitable for estimating the impact degree distributions.
In particular, the theoretical distributions obtained using the empirical $F(s)$ are narrower than the observed distributions although the theoretical distributions calculated using the estimated $F(s)$ are in good agreement with the observed distributions.
This result implies the existence of a synergistic effect of multiple knockouts on impact spreading in metabolic networks (i.e., the effect of multiple knockouts is larger than expected based on the assumptions regarding the combined effects of independent impact spreading initiated by individual knockouts).

\section{Discussion and conclusion}
\label{sec:discuss}
By extending the branching process approximation in the previous study \cite{Takemoto2012}, we proposed a random family forest model for estimating the impact degree distributions obtained as a result of {\it multiple} knockouts of metabolic networks, and demonstrated its validity using real-world metabolic network data.
Independent of our previous study \cite{Takemoto2012}, Lee et al. \cite{lee12} have also proposed a branching process approach for Boolean bipartite networks of metabolic reactions; however, they only considered cases including a single knockout.

In addition, we have provided a better definition of the number of offspring of a reaction node using the concept of spreading edges (see Sec. \ref{sec:offspring}).
Because of this modified definition, the poor predictions reported in our previous study \cite{Takemoto2012} were improved here in the context of a single knockout (see Table \ref{table:accuracy}).

However, the definition of spreading edges is not without limitations.
In particular, spreading edges do not allow for the consideration of conditional impact spreading.
For example, in Fig. \ref{fig2:reaction_net}B, the impact can spread from the reaction node $R_1$ to the reaction node $R_3$, if the reaction node $R_2$ has been already inactive; however, this definition is not applicable to such a situation.

The branching process approximation (i.e., the estimation using Eq. (\ref{eq:P(r)_emp})) also possesses limitations, such as the assumption of network tree (or less modular) structures, as explained in the previous study \cite{Takemoto2012}.
The existence of cycles may lead to an overestimation of the number of offspring for each reaction node because some offspring of a progenitor (reaction node) may have already been inactivated due to cycle structures; thus, the branching process approximation may overestimate the impact degree distributions.
Such overestimation of the impact degree distributions is observed in the case of multiple knockouts because the error is amplified through a power of the probability generating function of the impact degree (i.e., Eq. (\ref{eq:P(r)_k})) (see the solid lines in Figs. \ref{fig:fig4_double_knockout}B and \ref{fig:fig5_triple_knockout}B).
On the other hand, however, the number of offspring may also be underestimated because the definition of spreading edges does not consider conditional impact spreading (see Sec. \ref{sec:offspring} for details).
The estimation of offspring number is influenced by these 2 effects; therefore, the difficulty in estimating offspring number is also a limitation of the branching process approximation.

Analysis based on random family forest (i.e., the estimation using Eq. (\ref{eq:F(s)_2})) also has limitations such as an overestimation and/or underestimation of the impact degree distributions.
For example, due to independence, random family forests do not consider the integration of impact spreading initiated by initial impacts on metabolic networks.
In such cases, the impact degree distribution may be overestimated.
In contrast, the impact degree distribution may also be underestimated because the modeling based on random family forests does not consider the synergetic effect due to multiple knockouts; both computational (e.g., \cite{deutscher08}) and experimental studies (e.g., \cite{Deutscher2006,Suthers2009}) have reported such synergetic effects associated with multiple knockouts, indicating that genes (or reactions) not essential for growth, as in the case of a single knockout, are identified as essential genes (or reactions) when multiple knockouts are considered.
This effect is slightly related to conditional impact spreading (see Sec. \ref{sec:offspring} for details).
In Fig. \ref{fig2:reaction_net}B, for example, the impact does not spread if either the reaction $R_1$ or the reaction $R_2$ is inactivated (i.e., in the case of a single knockout).
However, the impact spreading occurs when both reactions $R_1$ and $R_2$ are inactivated (i.e., in the case of double knockouts).
Thus, the underestimation of the impact degree distributions (see the dashed lines in Figs. \ref{fig:fig4_double_knockout}C and \ref{fig:fig5_triple_knockout}C) might be observed.
The estimation of the impact degree distribution obtained in cases of multiple knockouts, when based on random family forests, is also complicated by these effects.

In order to make better predictions, analysis based on branching process may be improved by considering additional assumptions other than static metabolic network structures
For example, it may be useful to take into account the assumption of variable propagations in the branching process \cite{Dobson2010}, in which the mean of offspring distributions differs at each propagation stage.
For this modification, we would need to numerically obtain the offspring distributions at each propagation stage after the initial impact.
Although this procedure is less advantageous from the viewpoint of computational costs, such an improvement may become an important topic in the future.

Although the analysis based on branching process approximation possesses several limitations, as explained above, its validity has been confirmed overall using real-world metabolic networks.
This may be the result of the randomness or neutrality in metabolic network structures.
Several analytical \cite{Holme2011,Takemoto2013} and theoretical studies \cite{Lee2012,Takemoto2012b} suggest that the structure of metabolic networks can be determined through simple evolutionary processes and it is less modular (or more random) than previously thought.
Our results support the validity of the application of branching processes to the estimation of impact degree distributions.

The theoretical estimation of impact degree distributions is helpful for both experimental and computational studies on multiple knockouts because the evaluation of the impact of multiple knockouts is expensive.
For example, it is difficult to numerically obtain the impact degree distributions by the knockout of more than 3 reactions, even when an efficient algorithm \cite{tamura11} is used. This is because of the combinatorial complexity; thus, we could only test cases in which of fewer than 4 reactions are simultaneously are knocked out.
The analysis based on branching processes can easily predict the impact degree distributions obtained as a result of multiple knockouts using either the structure of metabolic networks or the impact degree distributions for single knockout, which can be easily calculated using the efficient algorithm \cite{tamura11}; however, some errors may be observed due to several limitations with increase in the number of disrupted reactions (i.e., in the case of high-order knockouts).
Such combinatorial complexity is also observed in experimental studies on multiple knockouts (e.g., \cite{Deutscher2006,Nakahigashi2009}).
In particular, it is hard to find the set of reactions significantly affecting metabolic dynamics.
Deutscher et al. \cite{deutscher08} also mentioned such a problem due to the combinatorial complexity in experimental studies.
Our theoretical approach may be able to support such experimental studies by using computational approaches, as introduced in Sec. \ref{sec:intro}, as well as studies of robustness in metabolic networks.

\section*{Acknowledgments}
This work was supported by a Grant-in-Aid for Young Scientists (A) from the Japan Society for the Promotion of Science (JSPS) (no. 25700030).
K.T. was partly supported by Chinese Academy of Sciences Fellowships for Young International Scientists (no. 2012Y1SB0014), and the International Young Scientists Program of the National Natural Science Foundation of China (no. 11250110508).
T.A. was partly supported by JSPS, Japan (Grants-in-Aid 22240009 and
22650045).



\end{document}